\documentclass[12pt]{iopart}

\usepackage[symbol]{footmisc}

\usepackage{graphicx}

\usepackage{soul,color}

\begin{document}

\title{Using HPC infrastructures for deep learning applications in fusion research}

\author{Diogo R. Ferreira$^1$ \& JET Contributors\footnote[1]{See the author list of E. Joffrin et al. 2019 Nucl. Fusion 59 112021.}}

\address{EUROfusion Consortium, JET, Culham Science Centre, Abingdon, OX14 3DB, UK}
\address{$^1$ Instituto de Plasmas e Fusão Nuclear (IPFN), Instituto Superior Técnico (IST), Universidade de Lisboa, 1049-001 Lisboa, Portugal}
\ead{diogo.ferreira@tecnico.ulisboa.pt}

\vspace{10pt}

\begin{indented}
\item[]Original version February 2021; Revised version May 2021
\end{indented}

\begin{abstract}
In the fusion community, the use of high performance computing (HPC) has been mostly dominated by heavy-duty plasma simulations, such as those based on particle-in-cell and gyrokinetic codes. However, there has been a growing interest in applying machine learning for knowledge discovery on top of large amounts of experimental data collected from fusion devices. In particular, deep learning models are especially hungry for accelerated hardware, such as graphics processing units (GPUs), and it is becoming more common to find those models competing for the same resources that are used by simulation codes, which can be either CPU- or GPU-bound. In this paper, we give examples of deep learning models -- such as convolutional neural networks, recurrent neural networks, and variational autoencoders -- that can be used for a variety of tasks, including image processing, disruption prediction, and anomaly detection on diagnostics data. In this context, we discuss how deep learning can go from using a single GPU on a single node to using multiple GPUs across multiple nodes in a large-scale HPC infrastructure.
\end{abstract}

%
%
%
%
%

\section{Introduction}

There is no question about the numerous advances that HPC-based modeling codes have introduced in the understanding of several plasma phenomena~\cite{goerler11gene,nardon16jorek,ku18xgc1}. In general, these are highly parallelizable codes that can simulate plasma behavior from first principles, under assumptions that are well grounded on magnetohydrodynamics and gyrokinetic theory~\cite{catto81gyrokinetics}. Most of these codes rely heavily on multiprocessing across as many CPUs as there are available, but in recent years there has been a growing interest in porting some of their routines to GPUs, with significant improvements in performance~\cite{azevedo13hybrid,snytnikov18pic}.

At the same time, there has been a growing interest in the application of machine learning~\cite{rea18exploratory} and deep learning~\cite{harbeck19predicting} to experimental data from fusion diagnostics, especially in connection with disruption prediction in tokamaks. In particular, the use of deep learning, which is fundamentally GPU-intensive, is bringing more researchers to HPC infrastructures and is increasing the demand for GPU-enabled clusters, such as the GPU partitions in the MARCONI-FUSION facility~\cite{iannone18marconi} or the Tiger GPU cluster at Princeton~\cite{churchill20deep}. In these and other HPC infrastructures, deep learning competes for computational resources with plasma modeling codes.

It is interesting to note that there are some fundamental differences in the type of computation that is performed by modeling codes and by deep learning. In essence, the aim of modeling codes is to investigate the physical phenomena associated with plasma behavior and, for this purpose, they use mostly synthetic data. In contrast, the aim of deep learning is to extract knowledge from the analysis of diagnostics data; in this context, experimental data from plasma diagnostics play the main role.

In addition, deep learning has a training stage and an inference stage, with the training stage being the most computationally intensive and time consuming, as a model is being trained to fit the data; in the inference stage, the model is used to make predictions, and ideally this should be as quick as possible. In simulation codes, the model is provided by theory rather than being learned from data, and running the model to make predictions is the single, most computationally heavy stage.

More recently, it has been found that deep learning can also be used to train surrogate models that can replace some of the most computationally expensive steps in a simulation~\cite{plassche20fast}. In this case, the surrogate model is trained to produce essentially the same results, with acceptable accuracy, while being much faster. This means that researchers working with modeling codes will soon be using deep learning for their own purposes, e.g.~by applying machine learning over the results of their simulations.

Either way, whether deep learning is applied to experimental data or to synthetic data, the use of deep learning in fusion research is bound to grow, and it will be more and more common to find deep learning tasks running alongside modeling codes on large-scale HPC infrastructures. Despite requiring its own set of software libraries, it is often possible to run deep learning tasks with relative ease on top of the software environment that is typically available in HPC infrastructures.

In this paper, we give examples of such deep learning tasks, namely:
\begin{itemize}
	\item training a convolutional neural network (CNN) for image reconstruction;
	\item training a recurrent neural network (RNN) for disruption prediction;
	\item training a variational autoencoder (VAE) for anomaly detection.
\end{itemize}

For simplicity, these examples will be based on a single plasma diagnostic, namely the JET bolometer diagnostic. This is a multi-channel diagnostic that provides a 2D view over a poloidal cross-section of the tokamak. The data from this diagnostic can be used either as a 1D vector of raw measurements, or as a 2D plasma profile after image reconstruction. Performing this image reconstruction is an example of a computationally intensive task that can be replaced by a deep learning model.

In section~\ref{sec:bolometer-diagnostic}, we provide an overview of the JET bolometer diagnostic, which will serve as the basis for the examples to be presented subsequently. In section~\ref{sec:plasma-tomography}, we describe a convolutional neural network for the reconstruction of the 2D plasma radiation profile from bolometer data; in section~\ref{sec:disruption-prediction}, we describe a recurrent neural network for disruption prediction from bolometer data; and in Section~\ref{sec:anomaly-detection}, we describe a variational autoencoder for anomaly detection on the 2D plasma profiles generated from bolometer data. All of these models have been trained on a single node with multiple GPUs, but at the end we also discuss the use of multiple nodes for hyperparameter tuning.

\section{The JET bolometer diagnostic}
\label{sec:bolometer-diagnostic}

The JET bolometer diagnostic~\cite{huber07upgraded} comprises two cameras -- a horizontal and a vertical camera -- that collect measurements of radiated power on a poloidal cross-section of the tokamak. Each camera has 24 sensors (bolometers), where each bolometer consists of a thin metal foil (about 10 $\mu$m) coupled with a temperature-sensitive resistance. As the plasma radiates power, the temperature of the metal foil fluctuates and there is a proportional change in the resistance, which can be measured to yield a linear response to the radiated power, in the ultraviolet to soft X-ray range~\cite{mast91bolometer}.

\begin{figure}[h]
	\centering
	\includegraphics[scale=0.6]{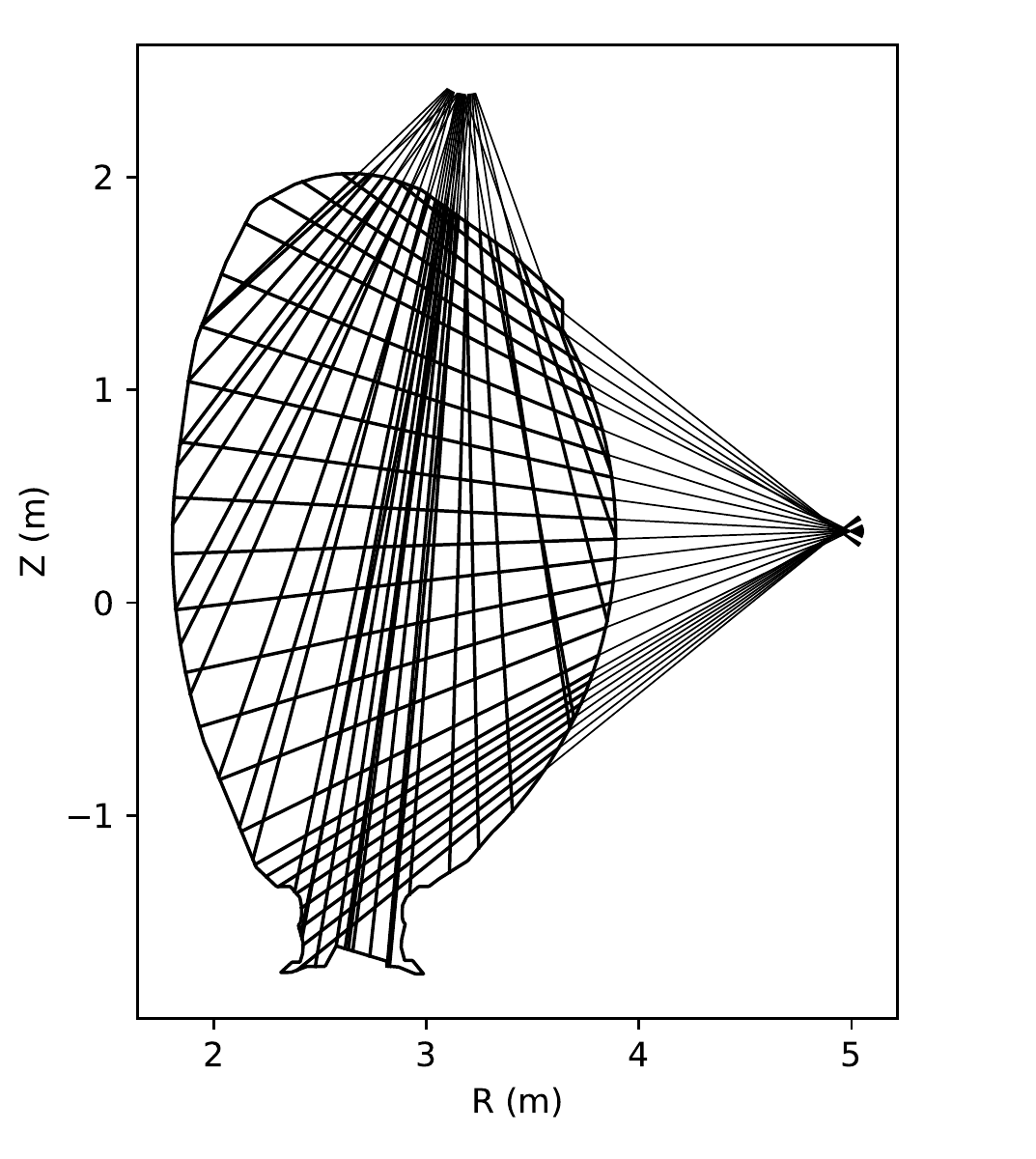}
	\caption{Lines of sight for the bolometer diagnostic at JET.}
	\label{fig_kb5}
\end{figure}

Each bolometer measures the line-integrated radiation along a particular line of sight. The geometrical arrangement of the lines of sight is such that, for each camera, 16 channels cover the whole plasma, and 8 channels provide a more fine-grained resolution over the divertor region. This arrangement is illustrated in figure~\ref{fig_kb5}.

From the bolometer measurements, it is possible to reconstruct the 2D plasma radiation profile using different tomography techniques~\cite{mlynar19current}. The underlying principle is the same as computed tomography (CT) in medical applications~\cite{buzug08book}, but the reconstruction method is different due to the scarce number of lines of sight available. The method that is used at JET uses an iterative constrained optimization algorithm that minimizes the error with respect to the bolometer measurements, while requiring the solution to be non-negative~\cite{ingesson98tomography}. Figure 2 shows a sample result.

\begin{figure}[h]
	\centering
	\includegraphics[scale=0.4]{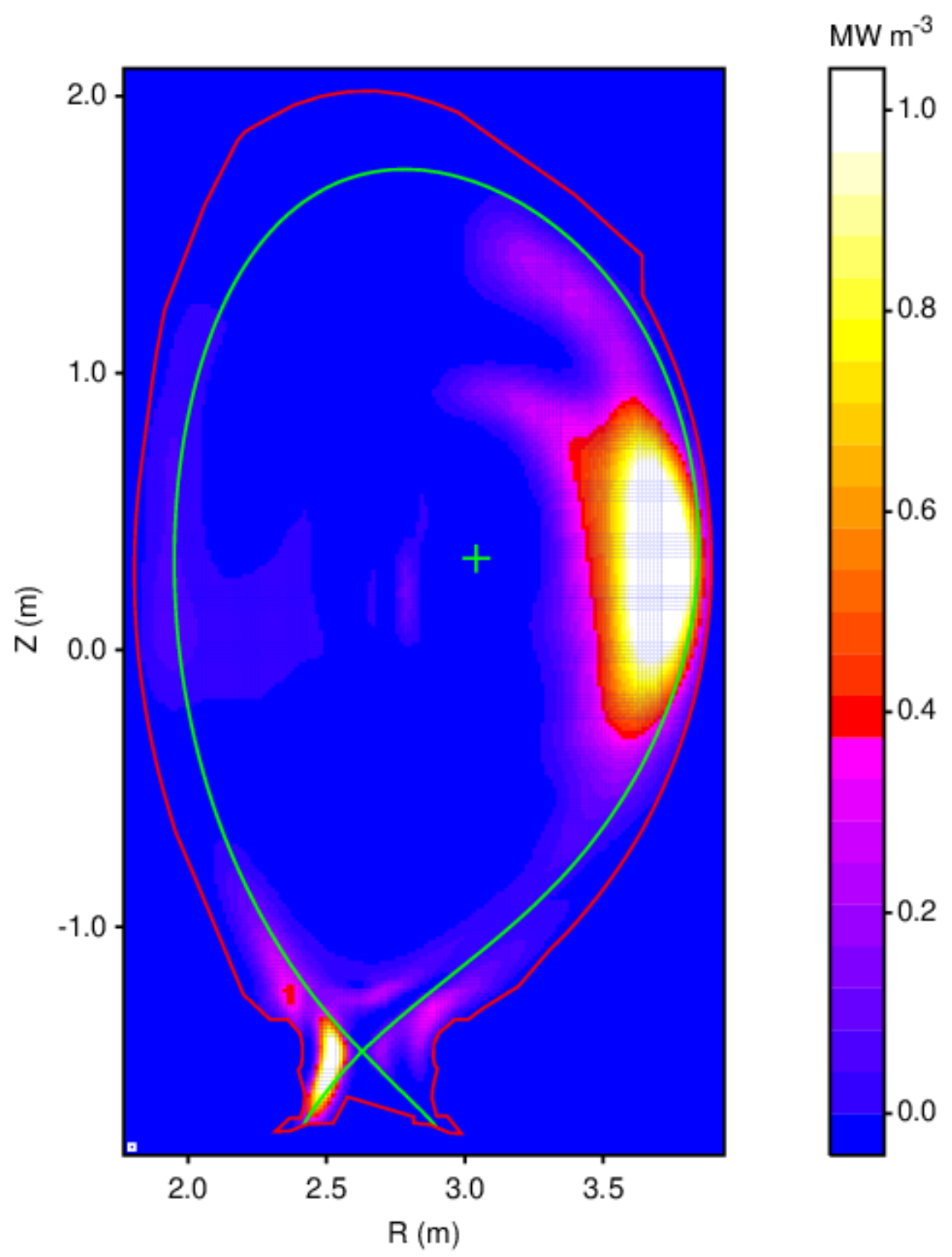}
	\caption{Example of a plasma radiation profile for JET pulse 92213 at $t$=50.0s.}
	\label{fig_tomo5}
\end{figure}

This iterative method takes a significant amount of computation time, typically on the order of minutes, to produce the plasma radiation profile. A deep neural network can generate the same results several orders of magnitude faster~\cite{ferreira18fullpulse}, at the cost of a small amount of error which, in any case, is below the experimental error associated with the input measurements ($\sim$10\%). This is the first example that we will describe in the next section. Subsequent examples will be based either on the bolometer measurements or the plasma profiles generated by this deep neural network.

\section{Deep learning for plasma tomography}
\label{sec:plasma-tomography}

In their most usual form, convolutional neural networks (CNNs) receive a 2D image as input and perform image classification through a series of convolutional and subsampling layers, followed by a couple of dense layers at the end~\cite{lecun15deep}. However, for the purpose of plasma tomography, it is more appropriate to have a network that takes a 1D vector of bolometer measurements as input and produces a 2D image of the plasma radiation profile as output. For this purpose, it is possible to reverse the structure of a typical CNN, by having a couple of dense layers at the beginning, and then a series of transposed convolutions or upsampling layers to produce an image as output~\cite{dosovitskiy17generate}.

\begin{figure}[h]
	\centering
	\includegraphics[scale=0.6]{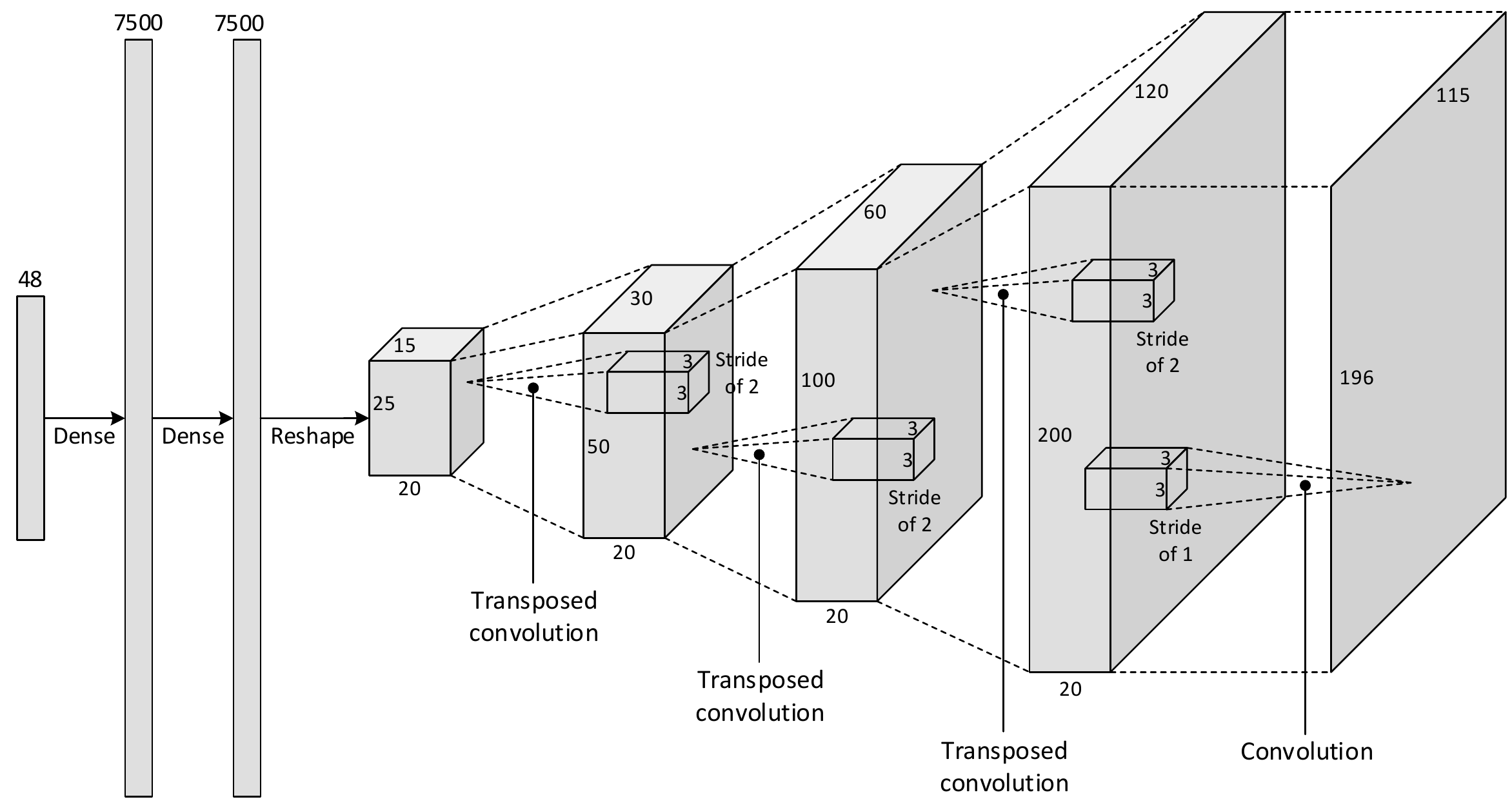}
	\caption{Structure of the deep learning model used for plasma tomography.}
	\label{fig_cnn}
\end{figure}

Figure~\ref{fig_cnn} shows the structure of the CNN used for plasma tomography. It receives the measurements from the bolometer diagnostic as input (48 lines of sight) and produces a plasma radiation profile at the output, with the same
resolution as the tomographic reconstructions that are routinely produced at JET (196$\times$115 pixels). The network comprises two dense layers with 7500 nodes, which are reshaped into a 3D tensor of size 25$\times$15$\times$20. This shape can be interpreted as comprising 20 features maps of size 25$\times$15. By applying a series of transposed convolutions, with activation functions in the form of rectified linear units (ReLUs), those feature maps are brought up to a size of 200$\times$120, from which the output image is generated by one last convolution.

The model was trained on about 28,000 sample reconstructions that had been pre-computed at JET across all the experimental campaigns since the installation of the ITER-like wall in 2011~\cite{matthews11ilw}. The dataset was split into 80\% for training, 10\% for validation, and 10\% for testing. To avoid overfitting, we used early stopping when the loss function (mean absolute error) on the validation set ceased to improve.

With a batch size of about 400 samples, it was possible to train the network on a single Nvidia P100 GPU, with 16 GB of memory. However, this took almost a day (16 hours) to train. In contrast, on a multi-GPU node with 8 Nvidia P100 GPUs, it was possible to train the model in a matter of hours (2-3 hours).

This illustrates the strong scaling the training process and highlights how beneficial it is to distribute the training across multiple GPUs, even when the training data could fit the memory of a single GPU by a proper choice of batch size. In terms of total training time, the improvement is roughly proportional to the number of GPUs available, and this can be achieved by changing only a few lines of code, when using recent deep learning frameworks such as TensorFlow~\cite{quanghung20tensorflow} or PyTorch~\cite{li20pytorch}.

Once trained, the model can be used to generate the reconstructions for an entire pulse in a matter of seconds (as opposed to taking several minutes to generate a single reconstruction, with the original method). The reconstruction error is of the order of 0.01 MW m$^{-3}$, which can be compared to the dynamic range of Figure~\ref{fig_tomo5} (1.0 MW m$^{-3}$) to conclude that the model provides good accuracy.

\begin{figure}[t]
	\centering
	\includegraphics[width=\textwidth]{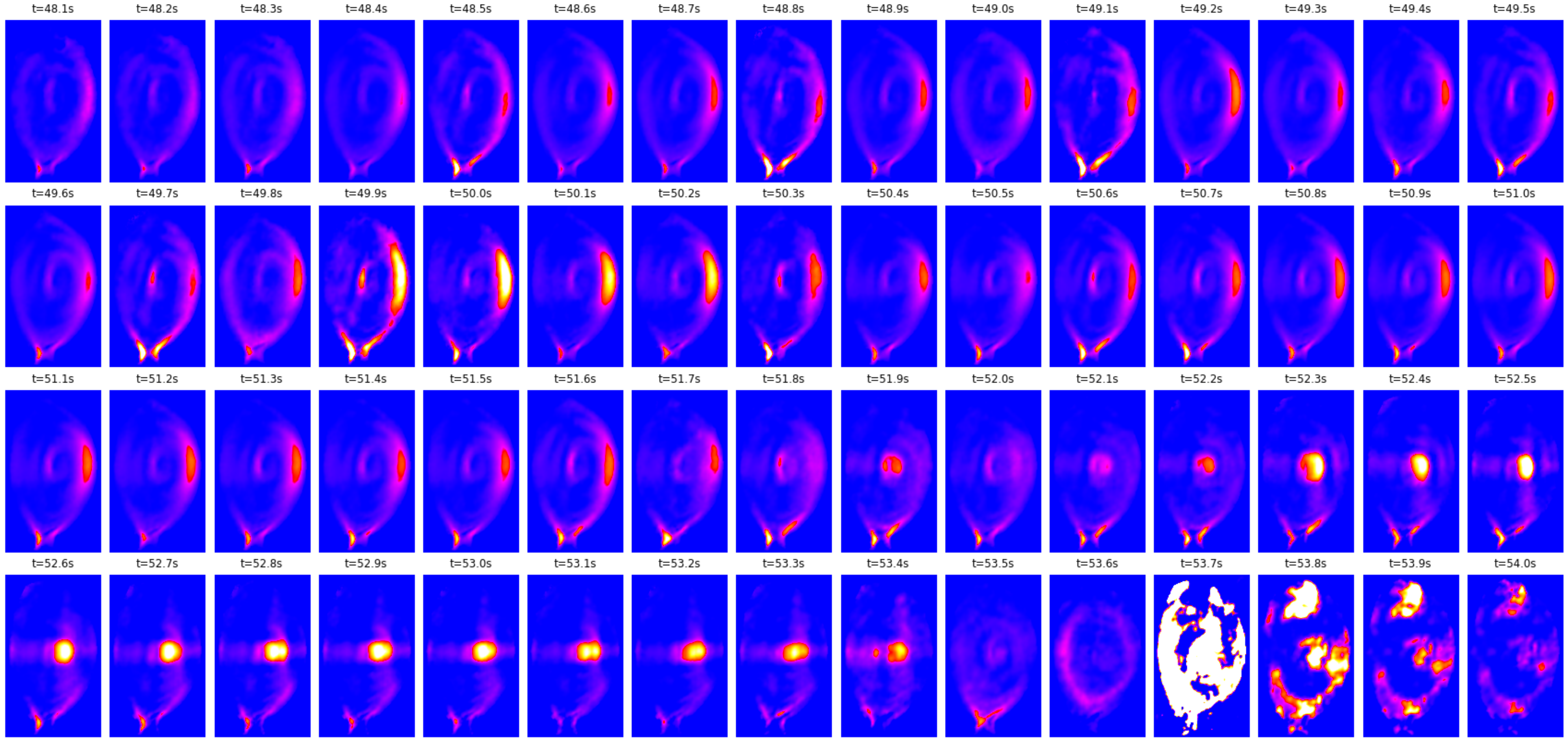}
	\caption{Reconstruction of pulse 92213 from 48.1s to 54.0s with a time step of 0.1s.}
	\label{fig_pulse}
\end{figure}

Figure~\ref{fig_pulse} shows an example of a full-pulse reconstruction, where pulse 92213 is being reconstructed from $t$=48.1s onwards, with a time step of 0.1s. Here it is possible to observe the development of a radiation blob at the outboard edge, followed by radiation peaking at the plasma core, which eventually leads to a disruption. This is in agreement with the literature, where one of the main reasons for disruption is reported to be impurity accumulation at the plasma core, leading to radiative collapse~\cite{joffrin13first}.

The fact that this behavior can be observed in the plasma radiation profile suggests that it should be possible to use the bolometer data for disruption prediction based on radiative phenomena. This brings us to the next example, which is to apply a recurrent neural network over the bolometer measurements in order to predict disruptions.

\section{Deep learning for disruption prediction}
\label{sec:disruption-prediction}

Over the years, a number of different approaches to disruption prediction have been developed, including the use of neural networks~\cite{cannas03neural}, support vector machines~\cite{cannas07svm} and decision trees~\cite{murari08trees}. More recently, the use of random forests~\cite{rea18exploratory} and deep learning~\cite{harbeck19predicting} have also shown promising results. Our goal is not to compete with those works, but to illustrate how feasible it is to develop a deep learning model for disruption prediction based on the JET bolometer diagnostic. The possibility of using this diagnostic for disruption prediction can be justified by the fact that, as illustrated in Figure~\ref{fig_pulse}, the plasma radiation profile provides important cues about disruption-relevant behavior.

In the previous section, we described a CNN that generates the plasma radiation profile from the bolometer measurements at a certain point in time. However, for the purpose of disruption prediction, it will be useful to consider how those measurements evolve over a period of time, in order to have a sense of how the plasma behavior is developing and how far it is from a possible disruption. Therefore, the input to the predictor should be a sequence of bolometer measurements, and this becomes an ideal scenario for the application of recurrent neural networks (RNNs).

\begin{figure}[t]
	\centering
	\includegraphics[scale=0.6]{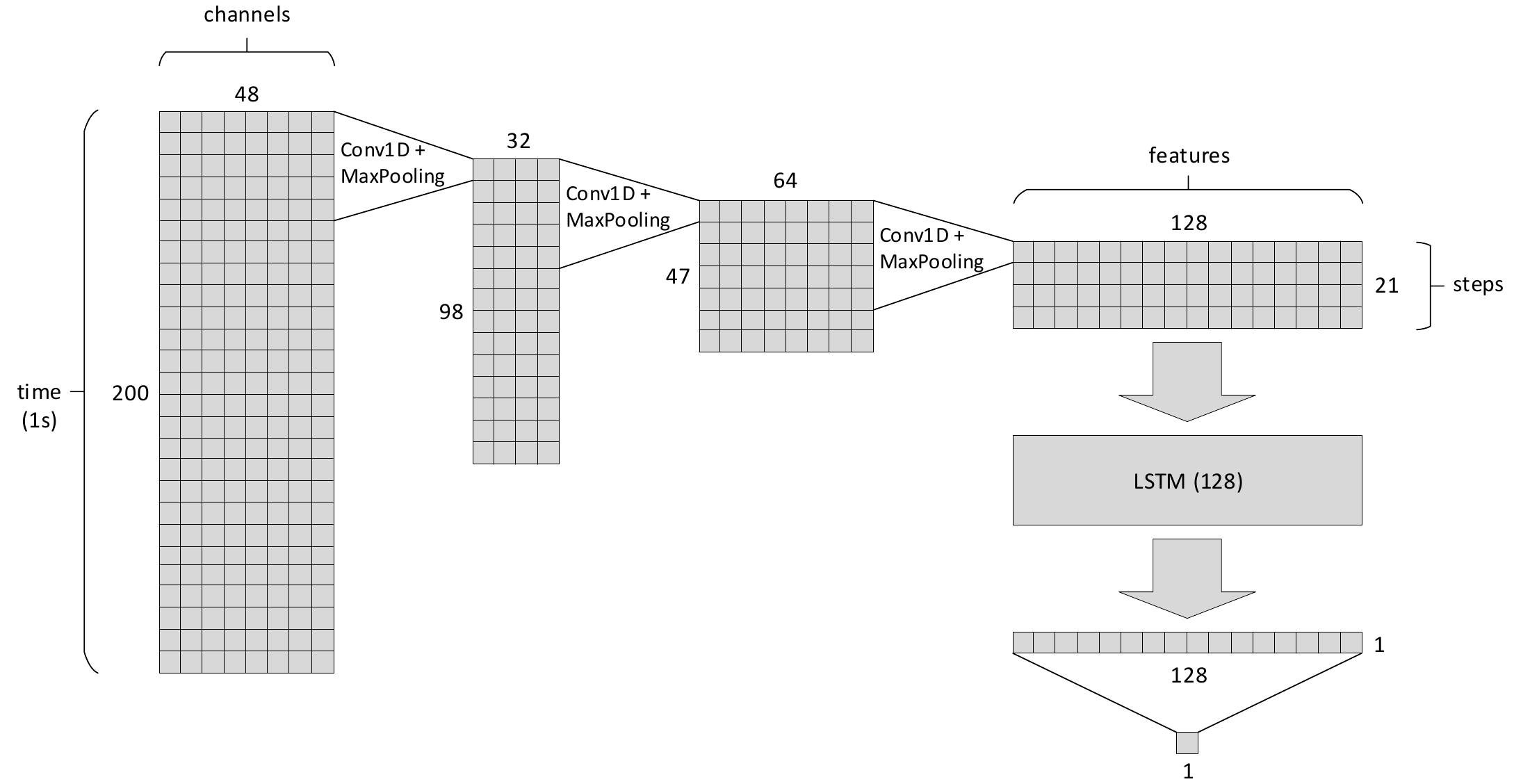}
	\caption{Structure of the deep learning model used for disruption prediction.}
	\label{fig_rnn}
\end{figure}

Figure~\ref{fig_rnn} shows a possible architecture for such network. In this case, the input has a shape of 200$\times$48, with the first being the sequence length (number of time steps), and the latter being the vector size (bolometer measurements). For the present purpose, the bolometer signals have been downsampled to 200 Hz, so a sequence length of 200 time points corresponds to exactly 1 second of bolometer data.

The model comprises a series of 1D convolutional layers that operate on the time axis. The use of subsampling (max-pooling) halves the sequence length after each convolution. On the other hand, the number of filters is doubled after each convolution, except on the first layer where the number of filters (32) is actually less than the number of input channels (48); this is done due to the redundancy and malfunctioning of certain channels. Eventually, this series of convolutional layers yields a sequence of 21 steps, where each step contains a 128-feature vector.

This sequence of feature vectors is handed over to an LSTM (long short-term memory)~\cite{hochreiter97lstm}, which returns the last output after having processed the entire sequence. One final dense layer combines the vector elements into a single output. In general, the structure of this network is similar to the type of CNN-LSTM models that are used for time series forecasting in many other fields, e.g.~\cite{livieris20forecasting}.

Depending on how the model is trained, the output can be either an estimated probability of disruption, or the time remaining to a predicted disruption. In fact, the model has been trained for both purposes, with the same architecture, but with two different prediction goals:
\begin{itemize}

	\item To train for probability of disruption, the training dataset was built by drawing random samples of bolometer data from about 10,000 pulses, including both disruptive and non-disruptive pulses, at any time point during those pulses. The samples were labeled as 1 or 0, depending on whether they came from a disruptive or non-disruptive pulse, respectively.

	\item To train for time to disruption, the training dataset was built by a similar procedure, but including disruptive pulses only, and only at time points before the disruption. Each sample was labeled with the time difference between the disruption time and the time point at which the sample was taken.

\end{itemize}

The two variants were trained simultaneously on separate GPUs. As before, although the model and the training data would fit the memory of a single GPU, using multiple GPUs made the training process significantly faster. However, rather than using a separate node to train each variant, we made a conservative use of resources by training both variants on the same multi-GPU node, with 4 GPUs allocated to training one variant, and 4 GPUs allocated to training the other.

In hindsight, such approach was probably not the most effective one, since the two variants have different training times, and the node is blocked until the longest of them completes. In this case, it would have been better to train the two variants using separate nodes. With twice the number of GPUs, the tasks would be twice as fast, and each node would be released as soon as the corresponding task completes.

\begin{figure}[h]
	\centering
	\includegraphics[scale=0.55]{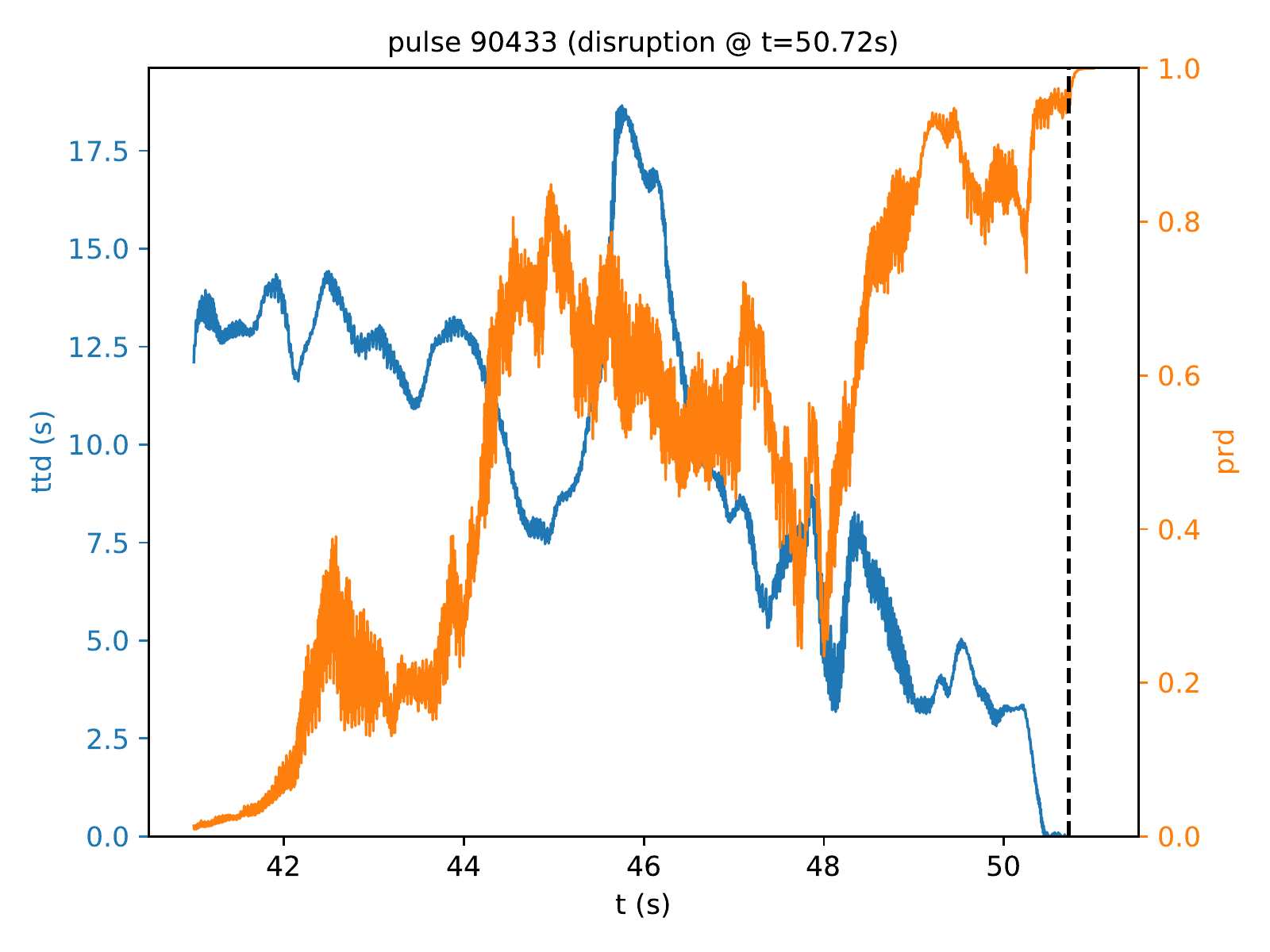}
	\caption{Time to disruption (\emph{ttd}) and probability of disruption (\emph{prd}) for a test pulse.}
	\label{fig_pred}
\end{figure}

Anyway, Figure~\ref{fig_pred} shows the predictions of the model for a test pulse, both in terms of estimated probability of disruption (\emph{prd}) and estimated time to disruption (\emph{ttd}), where the disruption that actually occurred is marked with a dashed vertical line. It is possible to observe that, a few hundred milliseconds before the disruption, the probability of disruption gets close do 1.0 and the time to disruption drops down to 0.0, both indicating that a disruption is imminent.

In fact, both indicators should be taken into account for a more accurate prediction but, even then, the success rate (recall) on a more extensive test set is only about 70\%~\cite{ferreira20prediction}, while other predictors currently being used at JET have a recall of about 85\%~\cite{moreno16prediction}. It should be noted, however, that these predictors are based on a wide range of global plasma parameters (e.g.~plasma current, locked mode amplitude, total input power, safety factor, poloidal beta, etc.) derived from several plasma diagnostics, which provide more information than the bolometer diagnostic alone.

Still, one can say that deep learning does seem to have the potential to find at least some disruption precursors in the bolometer data, and this provides the motivation for the next example, where we use anomaly detection to identify those precursors.

\section{Deep learning for anomaly detection}
\label{sec:anomaly-detection}

With a CNN, we were able to generate the plasma radiation profile at any point in time during a pulse; and with an RNN, we showed that there is relevant information for disruption prediction to be found in the bolometer data. The question now is whether it is possible to identify disruption precursors in the plasma profiles that are generated from bolometer data. In other words, if we look at the evolution of the plasma profile as in figure~\ref{fig_pulse}, it should be possible to say where exactly the disruptive behavior starts to appear. Our goal is to find this by automated means, using deep learning.

The problem falls within the scope of anomaly detection~\cite{zheng20anomaly,farias20automatic}, where a model is trained to tell whether a given sample is anomalous or not, based on how distinct that sample is from the samples that the model has seen during training. In our context, this implies training a model on non-disruptive pulses (the ``normal'' behavior), and then applying it to disruptive pulses in order to identify the anomalous behavior. If we run such model through an entire pulse, we should be able to see where the anomalies are.

In the field of deep learning, the type of model that is most popularly used for anomaly detection is the variational autoencoder (VAE)~\cite{kingma14autoenc}. In this case, anomaly detection relies on the ability of a VAE to reconstruct a given input sample. When the VAE struggles to reproduce the input sample, this is an indication that it has not seen such behavior before, and therefore such behavior is classified as an anomaly.

To reconstruct an input sample, the VAE produces an output that is of the same type and shape as its input. The difference (or error) between input and output is then calculated by some metric (or loss function), and the VAE is trained by minimizing such metric on the training set. The distinctive feature of the VAE is that it compresses its input into a probability distribution, and then it generates an output by sampling from such distribution. As a result, the output is a stochastic approximation of the input, but such approximation should be sufficiently good for the training data that the model has already seen; it is only for data that the model has not seen before that the reconstruction error will be large.

\begin{figure}[t]
	\centering
	\includegraphics[width=\textwidth]{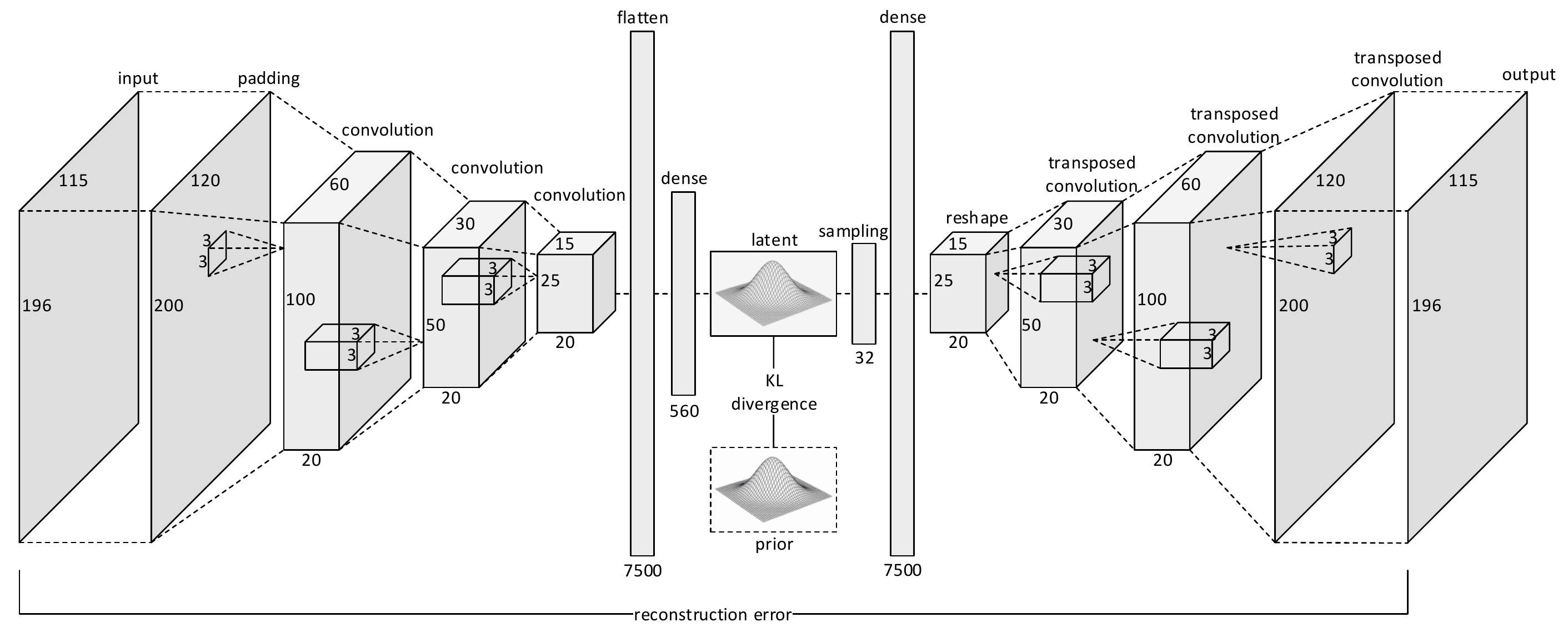}
	\caption{Structure of the deep learning model used for anomaly detection.}
	\label{fig_vae}
\end{figure}

Figure~\ref{fig_vae} shows the structure of a VAE that was designed for anomaly detection based on the plasma radiation profiles, such as those illustrated in Figure~\ref{fig_pulse}. For this purpose, the VAE receives a plasma profile as input, and produces another plasma profile as output. The reconstruction error (anomaly score) is defined as the mean absolute error per pixel between the input and output profiles. To compute the anomaly score across an entire pulse, one can provide the VAE with the plasma profiles that are generated by the model in Figure~\ref{fig_cnn}.

The advantage of performing anomaly detection on the plasma profile, rather than on the raw bolometer data, is that the plasma profile is already a preprocessed representation that eliminates some problems -- such as noise, malfunctioning sensors, and spurious measurements -- which could also be interpreted as anomalies.

As illustrated in Figure~\ref{fig_vae}, the VAE receives the plasma profile and includes a series of convolutional layers followed by a couple of dense layers that transform the input profile into a compressed representation. This representation is reinterpreted as providing the parameters of a multivariate normal distribution. A sampling layer draws a sample from this multivariate normal, and then a series of transposed convolutions transform that sample into an output profile.

The VAE therefore compresses (encodes) and decompresses (decodes) the plasma profile through a (latent) multivariate normal distribution. To prevent this distribution from overfitting the training data, it is often necessary to impose a prior, such that the latent distribution does not deviate too much from, say, a standard multivariate normal with zero mean and unit variance. This results in an additional (regularization) term in the loss function, based on the Kullback–Leibler divergence between the latent distribution and its prior. This additional term has effect during training only, and does not apply when the VAE is being run on test data.

The model was trained on about 1.4 million profiles generated from the bolometer data of about 250 non-disruptive pulses. Due to the amount of training data, the model was trained on a multi-GPU node using all of its 8 GPUs, taking about 15 hours to complete. Training the model on multiple nodes was not attempted, due to the difficulty of allocating nodes in a multi-user environment, and to hardware differences between nodes. Even though it would require changing only a few lines of code, the expected speedup did not seem to compensate the time waiting for resources to become available.

\begin{figure}[t]
	\centering
	\includegraphics[width=\textwidth]{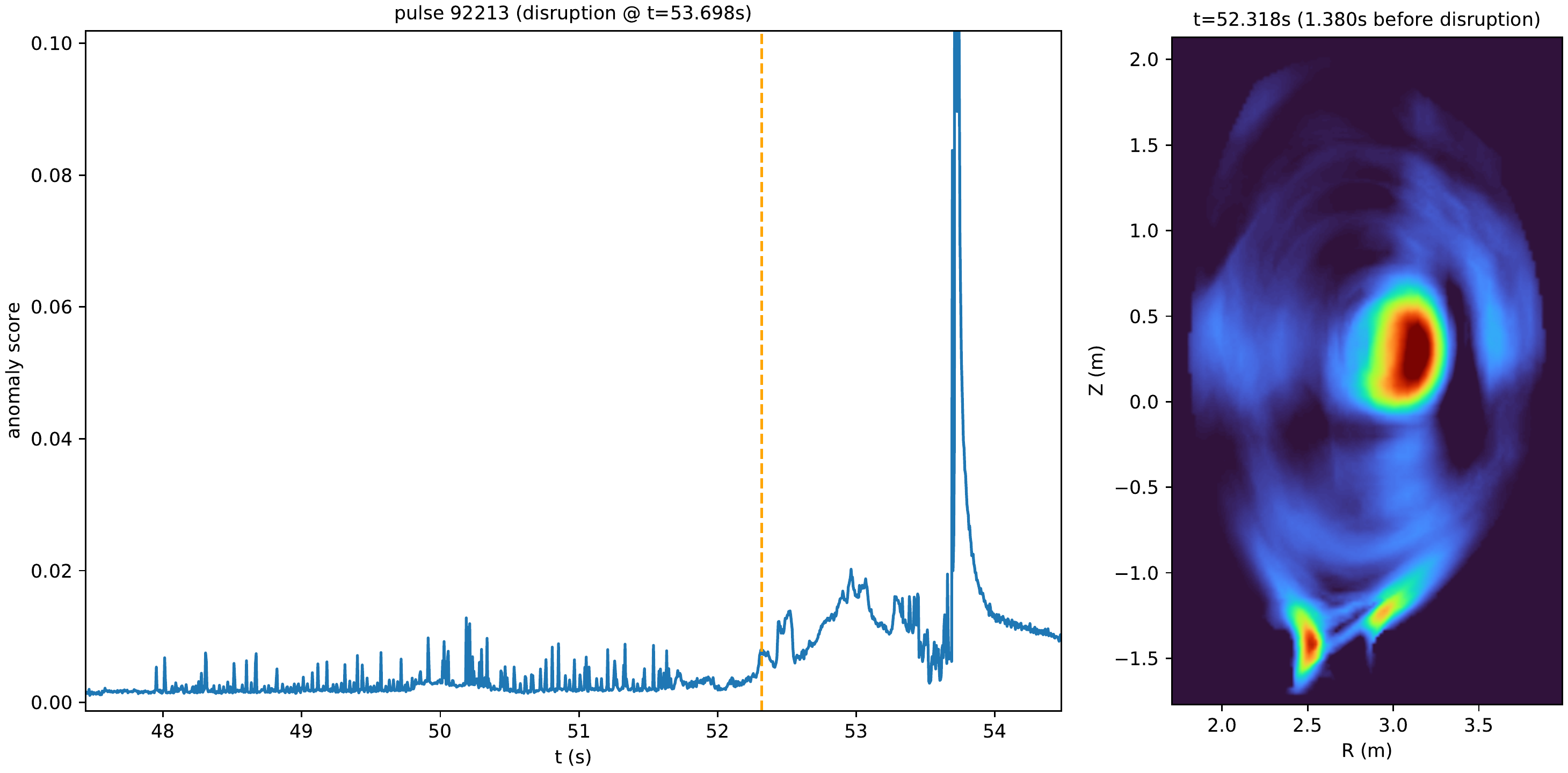}
	\caption{Anomaly score for pulse 92213 and plasma profile at the time indicated by the vertical dashed line (around $t$=52.3s).}
	\label{fig_anomaly}
\end{figure}

Having been trained on non-disruptive pulses, the model was then applied on a series of disruptive pulses, of which we report here only a single example. Figure~\ref{fig_anomaly} shows the reconstruction loss (anomaly score) provided by the VAE for the same pulse as in figure~\ref{fig_pulse}. As can be seen in figure~\ref{fig_anomaly}, there is a point at which the anomaly score begins to rise and stays relatively high for an extended period of time before the disruption. This corresponds to the moment when strong core radiation begins to develop and a radiation blob establishes itself into the core (cf.~figure~\ref{fig_pulse}). Multiple runs of the training procedure have yielded consistently the same results in this and other examples.

Although these results do not point to any new phenomena that has not already been discussed in the literature, they do strengthen the belief that strong core radiation, caused by impurity accumulation at the plasma center, is a characteristic behavior of disruptive pulses. As we can see from this experiment with a VAE, a model that has been trained on plasma profiles from non-disruptive pulses will assign a high anomaly score to profiles from disruptive pulses exhibiting strong core radiation. The same anomaly detection approach can be used to find other disruption precursors as well~\cite{ferreira20precursors}.

\section{Conclusion}

The availability of GPU partitions in most HPC clusters makes it possible to develop interesting applications of deep learning techniques, involving the analysis and processing of experimental data from plasma diagnostics, or of synthetic data produced by modeling codes. In most applications, it will be possible to train a deep learning model on a single node with multiple GPUs, where the number of GPUs available will provide a proportional speedup in the training process.

Most HPC infrastructures provide GPU nodes with 4 to 8 GPUs and an amount of RAM, typically 256 GB or more, which already enables a wide range of applications. In general, it is desirable to fit all training data into RAM to avoid any storage I/O during training. As for GPU memory, which is often the main bottleneck, this limit can often be circumvented with a proper choice of batch size. The choice of batch size is also important to ensure an efficient batching process between CPU and GPUs, in order to achieve maximum GPU utilization.

When doing hyperparameter tuning, for example by refining the model architecture or by choosing the best optimizer and learning rates, it may be necessary to run the training process multiple times with different setups. In this case, it becomes useful to run different setups on separate nodes, using at least two nodes to keep improving the model by pairwise comparison between alternative versions. Some hyperparameters, namely the number of layers and layer sizes, can also be found by taking into account the learning capacity required for a given problem or dataset.

As we see increasingly more powerful workstations being designed for deep learning, such as the Nvidia DGX Station A100, we also see that they are becoming prohibitively expensive,  driving more researchers to the use of shared HPC facilities, where they can leverage the economy of scale that results from aggregating the computational demands of many different projects. In this respect, the fusion community, with both its modeling and machine learning counterparts, is a paradigmatic example. They have more to gain by joining their computational demands, rather than keeping them apart.

\section*{Acknowledgments}

\footnotesize This work has been carried out within the framework of the EUROfusion Consortium and has received funding from the Euratom research and training programme 2014-2018 and 2019-2020 under grant agreement No 633053.
IPFN (Instituto de Plasmas e Fusão Nuclear) received financial support from FCT (Fundação para a Ciência e a Tecnologia) through projects UIDB/50010/2020 and UIDP/50010/2020. The authors are thankful for the granted use of computational resources provided by CCFE/UKAEA at Culham, UK, and by the MARCONI-FUSION HPC facility at CINECA, Italy.

\section*{References}

\bibliographystyle{iopart-num}
\bibliography{paper}

\end{document}